%% file: main.tex
\input{preamble}
\input{format}
\input{commands}

\begin{document}

\begin{Large}
    \textsf{\textbf{Listening with Language Models: Using LLMs to Collect and Interpret Classroom Feedback}}
    
\end{Large}

\vspace{1ex}

\textsf{\textbf{Sai Siddartha Maram:}} \text{University of California, Santa Cruz}, \href{mailto:samaram@ucsc.edu}{\texttt{samaram@ucsc.edu}}\\
\textsf{\textbf{Ulia Zaman:}} \text{University of California, Santa Cruz}, \href{mailto:uzaman@ucsc.edu}{\texttt{uzaman@ucsc.edu}}\\
\textsf{\textbf{Magy Seif El-Nasr:}} \text{University of California, Santa Cruz}, \href{mailto:mseifeln@ucsc.edu}{\texttt{mseifeln@ucsc.edu}}

\vspace{2ex}
\begin{abstract}
    Traditional end-of-quarter surveys often fail to provide instructors with timely, detailed, and actionable feedback about their teaching. In this paper, we explore how Large Language Model (LLM)-powered chatbots can reimagine the classroom feedback process by engaging students in reflective, conversational dialogues. Through the design and deployment of a three-part system—PromptDesigner, FeedbackCollector, and FeedbackAnalyzer—we conducted a pilot study across two graduate courses at UC Santa Cruz. Our findings suggest that LLM-based feedback systems offer richer insights, greater contextual relevance, and higher engagement compared to standard survey tools. Instructors valued the system’s adaptability, specificity, and ability to support mid-course adjustments, while students appreciated the conversational format and opportunity for elaboration. We conclude by discussing the design implications of using AI to facilitate more meaningful and responsive feedback in higher education.
\end{abstract}

\section{Introduction}

\begin{quote}
    \textit{Amanda teaches introductory Computer Science at UCSC. Each quarter, she awaits student feedback—hoping to understand what worked, what didn’t, and how she might improve. But all she gets is a set of generic end-of-quarter surveys filled with bar graphs, vague comments, and often, silence. Students rush through the questions, barely reading them, and this rarely reflects the realities of her course. Amanda is left guessing—Did her new lab structure help?  How did struggling students feel about the pacing?}
\end{quote}

Amanda’s story is not uncommon. Despite best intentions, traditional classroom feedback mechanisms, especially end-of-quarter surveys—often fail to capture the richness, nuance, and immediacy of students’ learning experiences \cite{gamlem2013student}. They are too late, too broad, and too detached from the specific contexts that faculty like Amanda care about. This disconnect raises a crucial design challenge: how might we improve the experience of giving feedback for students, while providing deeper, more actionable insight for faculty?

In this paper, we explore the potential of Large Language Model (LLM)-powered conversational agents to reimagine the classroom feedback process. By engaging students in reflective, conversational dialogues, LLM-based chatbots can elicit richer, more personalized insights—at scale. Unlike static surveys, these systems adapt to the context of the course, follow up on partial answers, and allow students to articulate their experiences in their own words. For instructors like Amanda, this means receiving not just numbers—but narratives. Through our work, we examine how such systems can bridge the disconnect between the needs of educators and the realities of student feedback.

\begin{figure}[!ht]
    \centering
    \includegraphics[width=\linewidth]{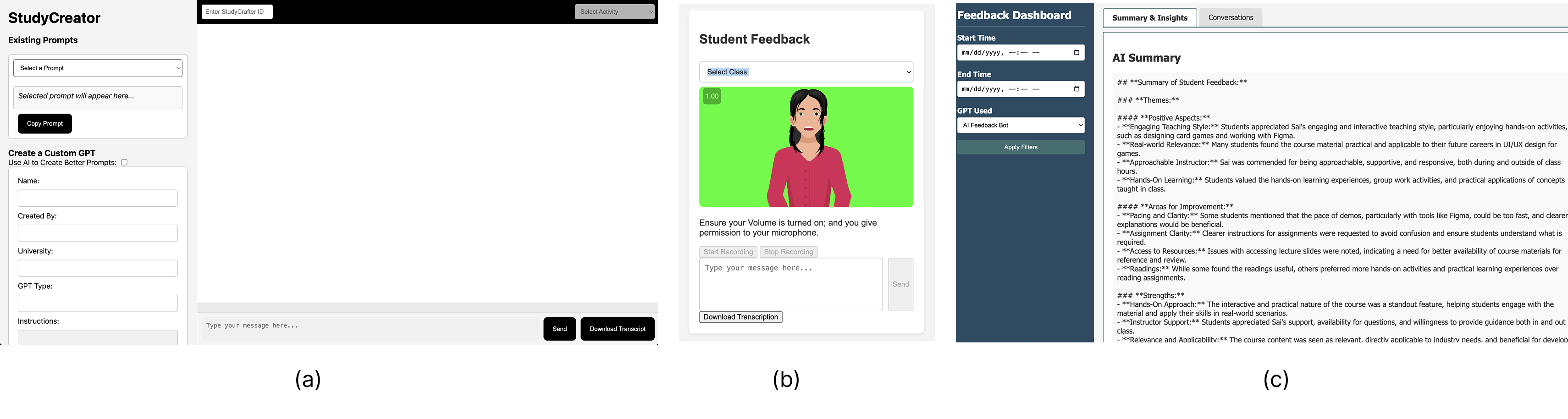}
    \caption{(a) The PromptCreator: A tool created for instructors to create and test their feedback prompts, (b) The FeedbackCollector: A student facing tool, where student can give their feedback, (c) The FeedbackAnalyzer: A tool for instructors, to view student responses, and gather AI summaries of the students responses. }
    \label{fig:the3systems}
\end{figure}

\section{Methodology}

To explore how LLM-based systems can enhance classroom feedback experiences, we employed a three-phase methodology involving prompt engineering via co-design, feedback interaction and collection, and insight synthesis and reflection. First, we collaborated with instructors using \textit{PromptDesigner}, a web-based tool we developed to help faculty craft, test, and refine conversational prompts tailored to their course needs. PromptDesigner enables instructors to upload syllabi, specify areas for feedback (e.g., assignment clarity, lecture engagement, classroom dynamics), and set parameters for timing and frequency of feedback requests. Two researchers facilitated participatory co-design sessions \cite{maram2024instructor,maram2024ah}, helping instructors align prompts with their pedagogical goals and classroom rhythms. 

Second, once the prompts were finalized, we deployed \textit{FeedbackCollector}, a student-facing chatbot powered by a large language model (GPT-4), to gather feedback throughout the quarter. Students accessed the chatbot asynchronously via a secure platform, engaging in conversational feedback sessions designed to encourage detailed reflections and clarifications. Third, after the quarter concluded, we conducted semi-structured interviews with six participating students and the instructor to reflect on their experiences using the chatbot. Faculty were additionally provided access to \textit{FeedbackAnalyzer}, a dashboard visualizing anonymized student-chatbot conversations, key themes, and conversational summaries. 

\section{Case Study}

As a pilot study, we deployed our LLM-based feedback system in two graduate classrooms at UC Santa Cruz—Game Data Science and Ethics in HCI—both taught by the same instructor and a total of 40 students. This setting allowed us to compare feedback dynamics across distinct course topics while maintaining instructional consistency. The instructor chose to use FeedbackCollector at two key points during the quarter: once at the 5-week mark and again at the 10-week mark. To support this, we conducted two co-design sessions using PromptDesigner: one in Week 3 and another in Week 8 to tailor prompts based on emerging classroom needs, incorporate specific feedback themes from the first round. At the end of the term, we conducted semi-structured interviews with six students and the instructor to evaluate their experiences with the chatbot, the depth and usefulness of the feedback received, and how it compared to the university’s standard end-of-quarter survey. With the transcripts of the semi-structured interview, we performed a close reading analysis and share our early insights \cite{pearson2010hci}.  While we acknowledge that our findings are based on only two classrooms and a small number of participants, and thus do not permit broad generalizations, our aim is to open a conversation within the educational research community around this emerging opportunity.

\section{Insights}
\subsection{Instructors Perspectives on Collecting Feedback using LLM-Systems}

The instructor expressed a strong preference for the LLM-based system over traditional end-of-quarter surveys (referred as SETS in UCSC), emphasizing how the chatbot offered a much richer, more actionable, and timely form of feedback. Unlike SETS, which rely heavily on broad, generalized questions and Likert-scale ratings, the LLM system allowed the instructor to ask course-specific questions that directly related to their teaching context. This flexibility was deeply valued: \textit{“As an instructor, you can create some specific questions depending on the course, depending on the lecture… it's interactive, you can ask a question, and the bot continues with the students to get more comments and feedback.”} The ability to tailor prompts to the curriculum created a feedback process that felt more relevant and grounded.

Efficiency also emerged as a significant benefit. Whereas SETS typically saw low participation rates and often required incentivization attempts like extra credit, the bot was quick to use and the conversation styled feedback made it engaging for students. \textit{“It takes five minutes, not even more than that. And then I got 23 feedbacks, which is way more then what I get in SETS}” The instructor also noted that collecting feedback mid-quarter—rather than waiting until the course's end—allowed for timely adjustments. This iterative process not only improved the instructor's teaching methods in real-time but also gave students a sense that their feedback was genuinely influencing the course.

Beyond ease of collection, the FeedbackAnalyzer helped the instructor uncover specific, actionable insights that SETS would have missed. For instance, feedback about a \textit{“misalignment between readings and lectures”} prompted a change in classroom discussions to incorporate more material from readings. The instructor appreciated gaining this nuanced view into student experiences: \textit{“It’s providing some more understanding of my teaching styles.”} Moreover, issues like communication confusion early in the quarter, which students might not have otherwise voiced, were surfaced through the bot, enabling mid-course corrections that students later recognized and appreciated.

The flexibility of updating the feedback prompts between collection points was another advantage the instructor valued highly. The instructor shared, \textit{“I created prompts more specifically related to their final project, which I didn’t ask at the beginning.”} This adaptability stood in stark contrast to the static nature of SETS, which the instructor described as poorly suited for dynamic classroom contexts where teaching strategies and student needs evolve throughout the quarter.

Finally, the instructor reflected on how engaging with the bot's conversational transcripts and summaries offered deeper insights into students themselves—not just into their experiences of the course. Reading the chats revealed broader patterns about student habits, expectations, and work ethic, such as the tendency for some students to avoid reading assignment instructions thoroughly or seeking frequent structured guidance. The instructor shares \textit{“The insights on how some specific instructions were not being noticed by students made me realize things needed to more clearer in my assignments. These kinds of relational insights about students would have been nearly impossible to glean from standardized survey responses."}

Overall, the instructor saw LLM-mediated feedback not just as a better alternative to SETS, but as a transformative tool for creating a more responsive, student-centered teaching experience.

\subsection{Students' Perspectives on Providing Feedback Using LLM Systems}

Students broadly welcomed the LLM-based feedback system. Across the interviews, students highlighted that the chatbot’s conversational format made providing feedback feel more reflective, meaningful, and less like a chore. For instance, S1 (student 1) shared that \textit{``this felt a bit unique. This felt a bit interesting. This felt a bit more conversational, which is a new thing that we have experienced. Other than that, it was still a prototype, but it worked actually pretty well... this was better because it went beyond MCQs, which was present in the conventional form of taking reviews from students, which was pretty much an upgrade.''}. Similarly, S2 appreciated how the bot encouraged elaboration rather than shallow responses, explaining, \textit{``Once the chatbot started, the conversation started, it was pretty smooth. Based on the responses, it did ask me follow-up questions to dive deeper and continue the conversation from there, which I felt was very genuine in how feedback should be. It prompted me to think more about it and gave a detailed feedback.''}.

The natural flow of the conversation emerged as a notable strength. Students felt that the chatbot’s follow-up questions helped sustain meaningful dialogue. S4 reflected that \textit{``the bot was kind of progressing in terms of the conversation. It wasn't just asking the same questions or even reframing the questions. It was coming up with different ones. It felt like a proper conversation, not repetitive, which was good to see.''}

The character and tone of the chatbot emerged as an area for growth. S1 proposed that a more informal, impartial persona could encourage greater candor: \textit{``If the character can be a bit more informal, then people would reply more, respond more freely, more candidly. It needs to be impartial—not like a TA or professor—so students can feel freer.''}. Issues around anonymity were significant to several students. S2 described feeling some initial hesitation: \textit{``It was a lot of time, it is a trigger that this will be anonymous. We didn't know how, obviously, but based on the trust that it'll be anonymized, I felt okay eventually. But if there was something more on the interface, maybe a little note saying, `Your responses will remain anonymous,' that would make it easier to be honest from the beginning.''}. S4 was even more direct, explaining, \textit{``For example, if I had some criticism about the course or the professor, I would definitely not put that in because I'm sure it would be read by the professor. If I had full confidence that it wouldn't be shared or that my identity wouldn’t be attached, I would've framed my feedback differently.''}. These concerns highlight the need for clearer, visible reassurances about privacy to foster full openness.



\section{Conclusion and Future Work}

Building on the promising early insights from our pilot study, we plan to extend this work by collaborating with a larger number of faculty across diverse disciplines and classroom settings. By deploying our systems in a wider range of courses, we aim to identify conversational feedback strategies, enhance adaptability to different pedagogical contexts, and address design challenges such as personalization, anonymity, and interface clarity. Future iterations will also incorporate iterative feedback from both instructors and students, allowing us to systematically evaluate how conversational feedback tools can meaningfully transform classroom feedback practices at scale.

\section{Acknowledgment}
The research team gratefully acknowledges the support of the Teaching and Learning Center at the University of California, Santa Cruz. We extend our sincere thanks to Jessie Dubreuil, Samara Foster, and Robin C Dunkin for their valuable support and contributions.

\bibliographystyle{plainnat}
\bibliography{references}

\end{document}

%% file: preamble.tex
\documentclass[letterpaper, 11pt]{extarticle}

\usepackage[english]{babel}
\usepackage[margin = 1in]{geometry}

\usepackage{mathrsfs}
\usepackage{amsfonts}
\usepackage{amsmath}
\usepackage{amsthm}
\usepackage{amssymb}
\usepackage{physics}
\usepackage{dsfont}
\usepackage{esint}

\usepackage{enumerate}
\usepackage[shortlabels]{enumitem}
\usepackage{framed}
\usepackage{csquotes}

\usepackage{float}
\usepackage{tabularx}
\usepackage{xcolor}
\usepackage{multicol}
\usepackage{subcaption}
\usepackage{caption}
\usepackage[round, authoryear]{natbib}

\usepackage{hyperref}
\definecolor{links}{rgb}{0.36,0.54,0.66}
\hypersetup{
   colorlinks = true,
    linkcolor = black,
     urlcolor = blue,
    citecolor = blue,
    filecolor = blue,
    pdfauthor = {Author},
     pdftitle = {Title},
   pdfsubject = {subject},
  pdfkeywords = {one, two},
  pdfproducer = {LaTeX},
   pdfcreator = {pdfLaTeX},
   }

%% file: format.tex
\usepackage{titlesec}
\usepackage[many]{tcolorbox}

\titlespacing*{\chapter}{0cm}{-2.0cm}{0.50cm}
\titlespacing*{\section}{0cm}{0.50cm}{0.25cm}

\newtcbtheorem[]{problem}{Problem}%
    {enhanced,
    colback = black!5, 
    colbacktitle = black!5,
    coltitle = black,
    boxrule = 0pt,
    frame hidden,
    borderline west = {0.5mm}{0.0mm}{black},
    fonttitle = \bfseries\sffamily,
    breakable,
    before skip = 3ex,
    after skip = 3ex
}{problem}

\tcbuselibrary{skins, breakable}


%% file: commands.tex
\makeatletter
\newcommand*\bigcdot{\mathpalette\bigcdot@{.5}}
\newcommand*\bigcdot@[2]{\mathbin{\vcenter{\hbox{\scalebox{#2}{$\m@th#1\bullet$}}}}}
\makeatother

